\documentclass[11pt,a4paper]{article}

\usepackage{amssymb}

\usepackage[dvips]{graphicx}

\begin{document}

\title{\bf Factorization of integrals, defining the $\beta$-function,
into integrals of total derivatives in $N=1$ SQED, regularized by
higher derivatives}

\author{K.V.Stepanyantz}

\maketitle

\begin{center}

{\em Moscow State University, physical faculty,\\
department of theoretical physics.\\
$119992$, Moscow, Russia}

\end{center}

\begin{abstract}
Some calculations in supersymmetric theories, made with the higher
derivative regularization, show that the $\beta$-function is given
by integrals of total derivatives. This is qualitatively explained
for the $N=1$ supersymmetric electrodynamics in all orders.
\end{abstract}

\unitlength=1cm


\section{Introduction}
\hspace{\parindent}

Quantum corrections in supersymmetric theories were studied for a
long time. In particular, a lot of papers were devoted to
calculations of a $\beta$-function and an anomalous dimension,
e.g. \cite{DRED}. Most calculations were made with the dimensional
reduction \cite{Siegel}, which is a special modification of the
dimensional regularization. The matter is that the dimensional
regularization \cite{tHooft} breaks the supersymmetry and is not
convenient for calculations in supersymmetric theories. However,
it is well known that the dimensional reduction is not
self-consistent \cite{Siegel2}. Ways, allowing to avoid such
problems, are discussed in the literature \cite{Stockinger}. Other
regularizations are also sometimes applied for calculations in
supersymmetric theories. For example, in Ref. \cite{Mas} the
two-loop $\beta$-function of the N=1 supersymmetric Yang--Mills
theory was calculated with the differential renormalization
\cite{DiffR}.

It is also possible to make calculations with the higher covariant
derivative regularization, proposed in Ref. \cite{Slavnov}. It was
generalized to the supersymmetric case in Ref.
\cite{Krivoshchekov} (another variant was proposed in Ref.
\cite{West_Paper}). This regularization is self-consistent and
does not break the supersymmetry. A calculation of the one-loop
$\beta$-function for the (non-supersymmetric) Yang--Mills theory,
regularized by higher covariant derivatives, was made in Ref.
\cite{Ruiz}. After corrections, made in subsequent papers
\cite{Asorey}, the result coincided with the well-known result,
obtained with the dimensional regularization \cite{Politzer}. It
was proved that in the one-loop approximation calculations with
the higher covariant derivative regularization always agree with
the results of calculations with the dimensional regularization
\cite{PhysLett}. Some calculations in the one-loop and two-loop
approximations were made for various theories
\cite{Rosten_Calc,Rosten_Int} with a variant of the higher
covariant derivative regularization, proposed in
\cite{Rosten_Reg}. The structure of the corresponding integrals
was discussed in Ref. \cite{Rosten_Int}.

A calculation of the three-loop $\beta$-function for the $N=1$
supersymmetric electrodynamics (SQED) regularized by higher
derivatives \cite{3LoopHEP} reveals an interesting feature of
quantum corrections: loop integrals, defining the
$\beta$-function, are integrals of total derivatives. (In Ref.
\cite{3LoopHEP} three-loop integrals, defining the
$\beta$-function, were reduced to integrals of total derivatives
using integration by parts.) A two-loop calculation, made with the
dimensional reduction in $N=1$ SQED and revealing a similar
feature, was presented in Ref. \cite{Smilga}, where the
factorization of integrands into total derivatives is explained in
all loops using a special technique, based on the covariant
Feynman rules in the background field method \cite{Grisaru}. Such
factorization allows to calculate one of the loop integrals
analytically and obtain the NSVZ $\beta$-function
\cite{NSVZ_Instanton}, which (in $N=1$ SQED) relates the
$\beta$-function in $n$-th loop with the anomalous dimension in
the previous loop. Due to this, application of the higher
covariant derivative regularization is sometimes very convenient
in the supersymmetric case. The fact that integrals, appearing
with the higher covariant derivative regularization, in the limit
of zero external momentum become integrals of total derivatives,
seems to be a general feature of supersymmetric theories. It was
verified in the two-loop approximation in \cite{PhysLettSUSY} for
a general renormalizable $N=1$ supersymmetric theory. A similar
results can be obtained with different versions of the higher
derivative regularization.

Using a usual supergraph technique for $N=1$ SQED, regularized by
higher derivatives, a partial explanation of the factorization of
integrands into total derivatives was made in \cite{SD} by
substituting a solution of the Ward identity into the
Schwinger--Dyson equation. Using this method it is possible to
extract a contribution, giving the exact NSVZ $\beta$-function in
all orders. However, there is also another contribution. In $N=1$
SQED this contribution is nontrivial starting from the three-loop
approximation. In the three-loop approximation it is given by an
integral of a total derivative and is equal to 0. It was
conjectured in \cite{SD} that this occurs in all loops. A partial
four-loop verification of this statement was made in Ref.
\cite{Pimenov}.

In this paper for $N=1$ SQED using usual supergraphs we try to
explain, why integrals, defining a $\beta$-function, are integrals
of total derivatives in all loops. Although the explanation is not
completely rigorous, we demonstrate that a key ingredient of the
proof is a special algebraic identity, obtained in this paper.
Moreover, the simple arguments, presented here, can be formulated
in a rigorous form, and the results can be compared with explicit
calculations. We will describe this in the forthcoming paper.

The paper is organized as follows:

In Sec. \ref{Section_SUSY} we recall basic information about $N=1$
SQED, regularized by higher covariant derivatives. In Sec.
\ref{Section_Three_Loop} we explain how the three-loop
$\beta$-function is related with the two-loop anomalous dimension
due to the factorization of integrands into total derivatives. A
qualitative explanation of this feature is given in Sec.
\ref{Section_Factorization} for $N=1$ SQED. The results are
briefly discussed in the Conclusion. In Appendix
\ref{Appendix_Three_Loop} we present explicit expressions for the
three-loop integrals, defining the $\beta$-function and the
anomalous dimension in $N=1$ SQED. An identity, which is a key
ingredient of our arguments, is proved in the Appendix
\ref{Appendix_Identity}.

\section{N=1 supersymmetric electrodynamics, regularized by
higher covariant derivatives}
\hspace{\parindent}\label{Section_SUSY}

Let us consider $N=1$ SQED, regularized by higher derivatives. In
the massless case it is described by the action\footnote{In our
notation $\eta_{\mu\nu}=\mbox{diag}(1,-1,-1,-1)$;\ \ $\theta^a
\equiv \theta_b C^{ba}$; $\theta_a$ and $\bar\theta_a$ denote the
right and left components of $\theta$, respectively.}

\begin{equation}\label{SQED_Action}
S = \frac{1}{4 e^2} \mbox{Re} \int d^4x\,d^2\theta\,W_a C^{ab}
R(\partial^2/\Lambda^2) W_b + \frac{1}{4}\int d^4x\, d^4\theta\,
\Big(\phi^* e^{2V}\phi + \widetilde\phi^* e^{-2V}\widetilde\phi
\Big),
\end{equation}

\noindent where $R(\partial^2/\Lambda^2)$ is a regulator, e.g. $R
= 1 + \partial^{2n}/\Lambda^{2n}$. This action is invariant under
the gauge transformations

\begin{equation}
\phi \to e^{i\Lambda}\phi;\qquad \widetilde\phi \to
e^{-i\Lambda}\widetilde\phi;\qquad V \to V + \frac{i}{2}(\Lambda^*
- \Lambda),
\end{equation}

\noindent where $\Lambda$ is an arbitrary chiral superfield. A
gauge is fixed by adding

\begin{equation}\label{Gauge_Fixing}
S_{\mbox{\scriptsize gf}} = - \frac{1}{64 e^2} \int
d^4x\,d^4\theta\, \Big(V D^2 \bar D^2 R(\partial^2/\Lambda^2) V +
V \bar D^2 D^2 R(\partial^2/\Lambda^2) V\Big)
\end{equation}

\noindent to the action. The term with higher derivatives does not
remove divergences in the one-loop approximation
\cite{Slavnov_Book}. In order to cancel the remaining one-loop
divergences, the Pauli-Villars determinants should be inserted
into the generating functional

\begin{equation} Z[J,j] = \int
D\mu\,\smash{\prod\limits_I} \Big(\det PV(V,M_I)\Big)^{c_I}
\exp\Big\{i (S + S_\Lambda + S_{\mbox{\scriptsize gf}} +
\mbox{Sources}) \Big\},
\end{equation}

\noindent where $\sum\limits_I c_I = 1$ and $\sum\limits_I c_I
M_I^2 = 0$. It is useful to present the Pauli--Villars
determinants in the following form:

\begin{equation}
\det PV(V,M_I) = \Big(\int D\Phi_I D\widetilde\Phi_I
e^{iS_{I}}\Big)^{-1},
\end{equation}

\noindent where $S_I$ is the action for the Pauli--Villars
fields,\footnote{This action differs from the one, used in
\cite{3LoopHEP}, because here the quotient of the coefficients in
the kinetic term and in the mass term does not contain the factor
$Z$. Using terminology of Ref. \cite{Arkani}, one can say that
here we calculate the canonical coupling $\alpha_c$, while in Ref.
\cite{3LoopHEP} the holomorphic coupling $\alpha_h$ was
calculated. Certainly, after the renormalization the effective
action does not depend on the definitions. However, the
definitions, used here, are much more convenient.}

\begin{equation}
S_{I} = \frac{1}{4} \int d^4x\,d^4\theta\,\Big(\Phi_I^* e^{2V}
\Phi_I + \widetilde\Phi_I^* e^{-2V} \widetilde\Phi_I\Big) +
\Big(\frac{1}{2}\int d^4x\,d^2\theta\,M_I \widetilde\Phi_I \Phi_I
+\mbox{h.c.}\Big).\nonumber
\end{equation}

\noindent We assume that $M_I = a_I \Lambda$, where $a_I$ are
constants. (Thus, there is the only dimensionful parameter
$\Lambda$ in the regularized theory.)

The generating functional for connected Green functions and the
effective action are defined by the standard way. Terms in the
effective action, corresponding to the renormalized two-point
Green function of the gauge superfield, have the form

\begin{equation}\label{D_Definition}
\Gamma^{(2)}_V = - \frac{1}{16\pi} \int
\frac{d^4p}{(2\pi)^4}\,d^4\theta\,V(-p,\theta)\,\partial^2\Pi_{1/2}
V(p,\theta)\, d^{-1}(\alpha,\mu/p),
\end{equation}

\noindent where $\alpha$ is a renormalized coupling constant, and

\begin{equation}
\partial^2\Pi_{1/2}  = -\frac{1}{8} D^a \bar D^2 D_a
\end{equation}

\noindent is a supersymmetric transversal projector. We calculate

\begin{equation}\label{We_Calculate}
\frac{d}{d\ln \Lambda}\,
\Big(d^{-1}(\alpha_0,\Lambda/p)-\alpha_0^{-1}\Big)\Big|_{p=0} = -
\frac{d\alpha_0^{-1}}{d\ln\Lambda} =
\frac{\beta(\alpha_0)}{\alpha_0^2},
\end{equation}

\noindent where $\alpha$ and $\Lambda$ are considered as
independent variables. The anomalous dimension is defined
similarly. If the two-point Green function of the matter
superfield in the massless limit is written as

\begin{equation}\label{Renormalized_Gamma_2}
\Gamma^{(2)}_\phi = \frac{1}{4} \int
\frac{d^4p}{(2\pi)^4}\,d^4\theta\,\Big(\phi^*(-p,\theta)\,\phi(p,\theta)
+\widetilde\phi^*(-p,\theta)\,\widetilde\phi(p,\theta)\Big)
\,(ZG)(\alpha,\mu/p),
\end{equation}

\noindent then the anomalous dimension is defined by

\begin{equation}
\gamma\Big(\alpha_0(\alpha,\Lambda/\mu)\Big) = -\frac{d}{d\ln
\Lambda} \Big(\ln Z(\alpha,\Lambda/\mu)\Big).
\end{equation}

\noindent The two-point Green functions are calculated using a
standard supergraph technique, described e.g. in textbooks
\cite{West,Buchbinder}.

\section{Three-loop calculation for SQED}
\hspace{\parindent}\label{Section_Three_Loop}

A calculation of the three-loop $\beta$-function gives the
following result:

\begin{eqnarray}\label{Relation}
&& \frac{d}{d\ln\Lambda} \Big(d^{-1}(\alpha_0,\Lambda/p) -
\alpha_0^{-1}\Big)\Big|_{p=0} = - \frac{d}{d\ln\Lambda}
\alpha_0^{-1}(\alpha,\mu/\Lambda) = \frac{\beta(\alpha_0)}{\alpha_0^2}
\nonumber\\
&& = \frac{1}{\pi}\Big(1-\frac{d}{d\ln\Lambda} \ln
G(\alpha_0,\Lambda/q)\Big|_{q=0}\Big) = \frac{1}{\pi}-
\frac{1}{\pi}\frac{d}{d\ln\Lambda} \Big(\ln ZG(\alpha,\mu/q) - \ln
Z(\alpha,\Lambda/\mu)
\Big)\Big|_{q=0} =\nonumber\\
&& = \frac{1}{\pi} \Big(1 -
\gamma\Big(\alpha_0(\alpha,\Lambda/\mu)\Big)\Big).\qquad
\end{eqnarray}

\noindent The explicit three-loop expression for the left hand
side and the two-loop expression for the right hand side are
presented in Appendix \ref{Appendix_Three_Loop}. It should be
noted that they are given by well defined integrals. In
particular, the higher derivative regularization and the
differentiation with respect to $\ln\Lambda$ ensure that there are
no IR divergences. This agrees with the results of Ref.
\cite{Fargnoli} that the IR region does not affect to the
$\beta$-function.

In the three-loop approximation the integrals, obtained with the
higher covariant derivative regularization, can not be calculated
analytically. However, it is possible to find relation
(\ref{Relation}) between them. It is easy to see that this
relation appears due to the factorization of integrands into total
derivatives.

\section{Factorization of integrands into total derivatives
in $N=1$ SQED: qualitative explanation}
\hspace{\parindent}\label{Section_Factorization}

Now let us try to explain the factorization of integrands into
total derivatives in $N=1$ SQED, described by action
(\ref{SQED_Action}). We will calculate the two-point Green
function of the gauge superfield. Due to the Ward identity this
function is transversal. In order to obtain the $\beta$-function
we make calculations in the limit $p\to 0$ after differentiation
with respect to $\ln\Lambda$:

\begin{equation}
\int d^4\theta\, V\partial^2\Pi_{1/2} V \times
\frac{d}{d\ln\Lambda}\mbox{(Momentum integral)}\Big|_{p=0}.
\end{equation}

\noindent Therefore, in order to find a $\beta$-function it is
possible to make the substitution

\begin{equation}
V \to \bar\theta^a\bar\theta_a \theta^b\theta_b\equiv \theta^4,
\end{equation}

\noindent so that

\begin{equation}
\int d^4\theta V\partial^2\Pi_{1/2} V \to -8.
\end{equation}

\noindent This is possible, because we make calculations in the
limit $p\to 0$, where $p$ is an external momentum. (In this limit
the gauge superfield $V$ does not depend on the coordinates
$x^\mu$.)

We will try to reduce a sum of Feynman diagrams for the considered
theory to integrals of total derivatives. In the coordinate
representation such an integral can be written as

\begin{equation}
\mbox{Tr} \Big([x^\mu, \mbox{Something}]\Big) = 0.
\end{equation}

\noindent In order to extract such commutators first we consider
diagrams, containing a vertex to which only one external line (and
no internal lines) is attached. We can sum such a diagram with the
diagram, in which the external line is shifted to the nearest
vertex \cite{Identity}. This gives a sum of the subdiagrams,
presented below. Making the substitution $V \to \theta^4$ we
obtain

\vspace*{5mm}

\begin{picture}(0,0)
\put(-0.5,-1.5){\includegraphics[scale=0.4]{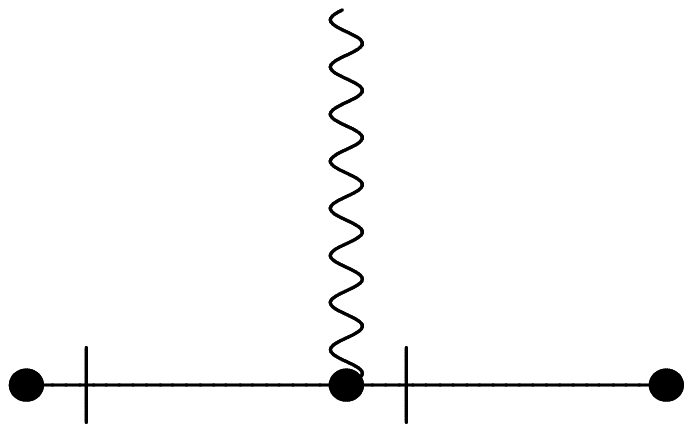}}

\put(3.1,-0.5){$+$}

\put(3.5,-1.5){\includegraphics[scale=0.4]{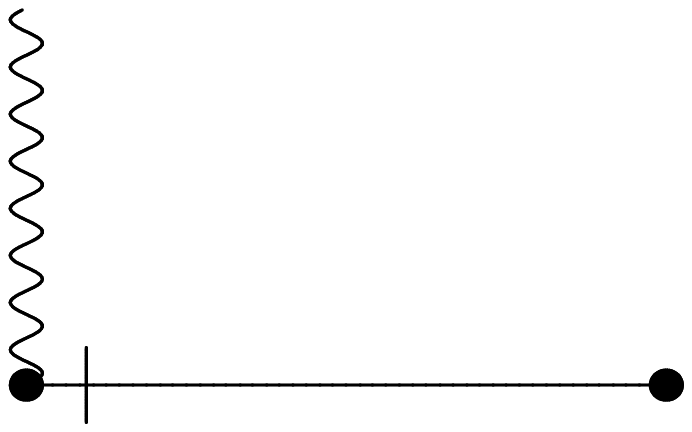}}

\put(7.2,-0.5){$=$}

\end{picture}

\vspace{8mm}
\begin{equation}
= -\theta^a \theta_a \bar\theta^b \frac{\bar D_b D^2}{4\partial^2}
+ \theta^a \theta_a \frac{D^2}{4\partial^2} + i \bar\theta^b
(\gamma^\mu)_b{}^a \theta_a \frac{\bar D^2 D^2
\partial_\mu}{8\partial^4} - i\theta^a
(\gamma^\mu)_a{}^b \frac{\bar D_b D^2
\partial_\mu}{4\partial^4} + \frac{\bar D^2 D^2}{16\partial^4}.
\end{equation}

\vspace*{5mm}

Only the first and the third terms give nontrivial contributions
to the two-point function of the gauge superfield, because they
contain $\bar\theta$. Really, finally it is necessary to obtain
$$
\int d^4\theta\,\theta^a \theta_a \bar\theta^b \bar\theta_b,
$$
while calculation of the supergraph can not increase degrees of
$\theta$ or $\bar\theta$. Therefore, we should have $\bar\theta^a
\bar\theta_a$ from the beginning.

Now let us proceed to calculation of Feynman diagrams. In the
one-loop approximation all calculations have been already done, so
that we can ignore this case. (In the one-loop approximation
contributions of the Pauli--Villars field are very essential, but
here we do not consider them. For investigation of the
Pauli--Villars contribution it is necessary to consider the
massive case. In principle, this is made similarly, but all
expressions become more complicated.)

First we consider diagrams, in which external $V$-lines are
attached to different loops of matter superfields. Let us denote a
chain of the matter superfield propagators, connecting vertexes
with quantum gauge field, by $*$.\footnote{It is possible to give
a rigorous definition of $*$. This will be done in
\cite{Future_Publication}. Below for simplicity we omit some
details, needed in the rigorous approach.} Then each loop will be
proportional to

\begin{eqnarray}
&& \mbox{Tr}\Big( i\bar\theta^c (\gamma^\nu)_c{}^d \theta_d
\frac{\bar D^2 D^2
\partial_\nu}{8\partial^4} * - \theta^c\theta_c \bar\theta^d
\frac{\bar D_d D^2}{4\partial^2} *\Big) \sim \mbox{Tr}\Big( -
\theta^c\theta_c \bar\theta^d
* \frac{\bar D_d D^2}{4\partial^2} *\nonumber\\
&& - \bar\theta^d \theta^c * \frac{\bar D^2 D_c}{4\partial^2} *
\frac{\bar D_d D^2}{4\partial^2} * + i\bar\theta^c
(\gamma^\nu)_c{}^d \theta_d * \frac{\bar D^2 D^2
\partial_\nu}{8\partial^4} *
+\mbox{$\theta^2$,$\bar\theta^1$,$\theta^1$,$\theta^0$
terms}\Big).
\end{eqnarray}

\noindent (Here we do not distinguish $*$ and $*^2$. This is
possible only qualitatively. However, the calculation can be
repeated using a rigorous approach. In this case the result should
be slightly modified.) After some simple algebra this expression
can be written as

\begin{equation}
\mbox{Tr}\Big( - 2\theta^c\theta_c \bar\theta^d [\bar\theta_d, * ]
+ i \bar\theta^c (\gamma^\nu)_c{}^d \theta_d
[y_\nu^*,*]+\ldots\Big) = 0 +\ldots,
\end{equation}

\noindent where $y_\nu^* = x_\nu - i\bar\theta^a
(\gamma_\nu)_a{}^b \theta_b$, and dots denote terms, proportional
to $\theta^2$, $\bar\theta^1$, $\theta^1$, $\theta^0$. In order to
calculate the two-point function we should multiply two such
expressions. Then the terms, denoted by dots, vanish due to the
integration over $d^4\theta$. Therefore, the sum of all such
diagrams is given by an integral of a total derivative and is
equal to 0.

Now let us consider diagrams, in which external $V$-lines are
attached to a single loop of the matter superfields.

We can shift $\theta$-s to an arbitrary point of the loop,
commuting them with matter propagators. Then we obtain

\vspace*{1.5cm}

\hspace*{4cm}
\begin{picture}(0,0)
\put(-4.5,0.1){\includegraphics[scale=0.4]{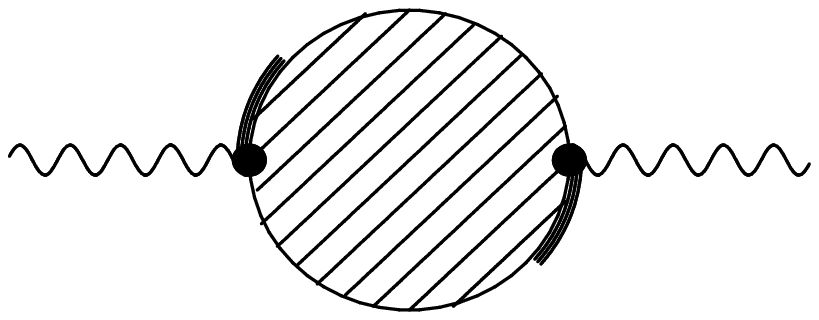}}
\put(-4.6,0.2){$\theta^a \bar\theta^b$}
\put(-1.8,0.2){$\theta^c\bar\theta^d$}
\put(-0.5,0.6){${\displaystyle \sim
\frac{i}{32}\frac{d}{d\ln\Lambda} \mbox{Tr}\Big( \theta^4
\frac{\bar D^2 D^2
\partial^\mu}{\partial^4} * \frac{\bar D^2 D^2
\partial_\mu}{\partial^4} *\Big);}$}

\put(-4.5,-1.6){\includegraphics[scale=0.4]{ins3.eps}}
\put(-4.6,-1.5){$\theta^a \bar\theta^b$}
\put(-2.1,-1.5){$\theta^c\theta_c\bar\theta^d$}

\put(-0.5,-1.1){${\displaystyle \sim 4 (\gamma^\mu)_d{}^c
\frac{d}{d\ln\Lambda} \mbox{Tr}\Big(\theta^4 \Big( \frac{\bar D^2
D_c \partial_\mu}{16\partial^4} * \frac{\bar D^d D^2}{\partial^2}
*} $}

\put(5.2,-2.3){${\displaystyle + \frac{\bar D^2 D^2
\partial_\mu}{16\partial^4} * \frac{\bar D^2 D_c}{16\partial^2} *
\frac{\bar D^d D^2}{\partial^2} *\Big)\Big);} $}

\put(-4.5,-4.3){\includegraphics[scale=0.4]{ins3.eps}}
\put(-4.6,-4.2){$\theta^a\theta_a \bar\theta^b$}
\put(-2.1,-4.2){$\theta^c\theta_c\bar\theta^d$}

\put(-0.5,-3.8){ $ {\displaystyle \sim -4i \frac{d}{d\ln\Lambda}
\mbox{Tr}\Big( \theta^4 \Big( - \frac{\bar D_d D^2}{4\partial^2}
* \frac{\bar D^2}{8\partial^2}
* \frac{\bar D^d D^2}{4\partial^2} * - \frac{\bar D_d}{2\partial^2}
* \frac{\bar D^d D^2}{4\partial^2} *}$}

\put(-0.5,-5.0){ $ {\displaystyle + \frac{\bar D_d
D^c}{2\partial^2} * \frac{\bar D^2 D_c}{8\partial^2} * \frac{\bar
D^d D^2}{4\partial^2} * + \frac{\bar D_d D^2}{4\partial^2}
* \frac{\bar D^2 D^c}{8\partial^2} * \frac{\bar D^2
D_c}{8\partial^2} * \frac{\bar D^d D^2}{4\partial^2} * \Big)\Big).
}$}
\end{picture}

\vspace*{5.8cm}

In order to write the sum of these diagrams as an integral of a
total derivative, we will start with the calculation of the
following sum:

\vspace*{1.3cm}

\begin{picture}(0,0)
\put(3.3,0.2){\includegraphics[scale=0.4]{ins3.eps}}
\put(3.3,0.3){$\theta^a \bar\theta^b$}
\put(5.9,0.3){$\theta^c\bar\theta^d$}

\put(7.5,0.7){$+$} \put(8.2,0.7){${\displaystyle\frac{1}{2}}$}

\put(9.3,0.2){\includegraphics[scale=0.4]{ins3.eps}}
\put(9.3,0.3){$\theta^a \bar\theta^b$}
\put(11.7,0.3){$\theta^c\theta_c\bar\theta^d$}

\end{picture}

Using the identity

\begin{equation}
[x^\mu,\frac{\partial_\mu}{\partial^4}] =
[-i\frac{\partial}{\partial p_\mu}, -\frac{ip_\mu}{p^4}] = -2\pi^2
\delta^4(p_E) = -2\pi i \delta^4(p)
\end{equation}

\noindent after some algebra we obtain that this sum is given by

\begin{equation}\label{Singularity}
4i \frac{d}{d\ln\Lambda} \mbox{Tr}\Big(\theta^4
\Big(\frac{i\pi^2}{8}
* \bar D^2 D^2 \delta^4(\partial_\alpha) +
\Big[y_\mu^*, \frac{\bar D^2 D^2\partial^\mu}{16\partial^4}
*\Big]\Big)\Big)  = -\mbox{Tr}\Big(\frac{\pi^2}{2} \theta^4
* \bar D^2 D^2 \delta^4(\partial_\alpha)\Big).
\end{equation}

$\delta$-function allows to perform integration over the momentum
of the considered matter loop. This corresponds to cutting the
diagram and gives diagrams for the two-point Green function of the
matter superfield \cite{Smilga}. For example,

\vspace*{1.2cm}
\begin{picture}(0,0)
\put(2.8,-0.5){\includegraphics[scale=0.24]{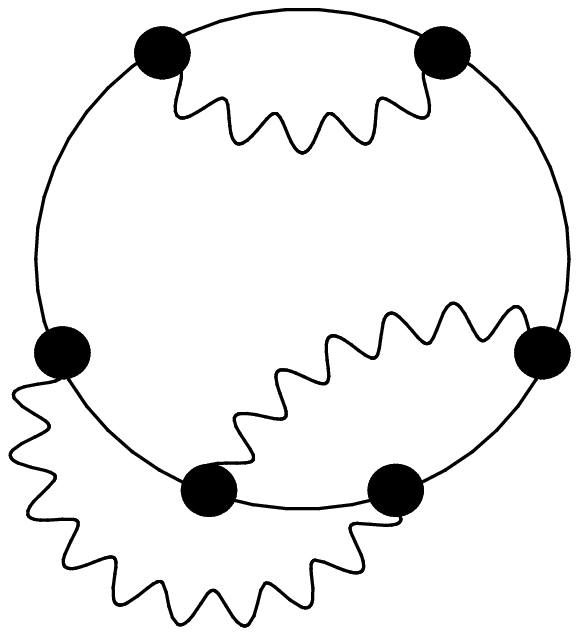}}
\put(3.5,0.3)

\put(5,0.6){\vector(1,0){1.4}}

\put(6.6,-0.5){\includegraphics[scale=0.24]{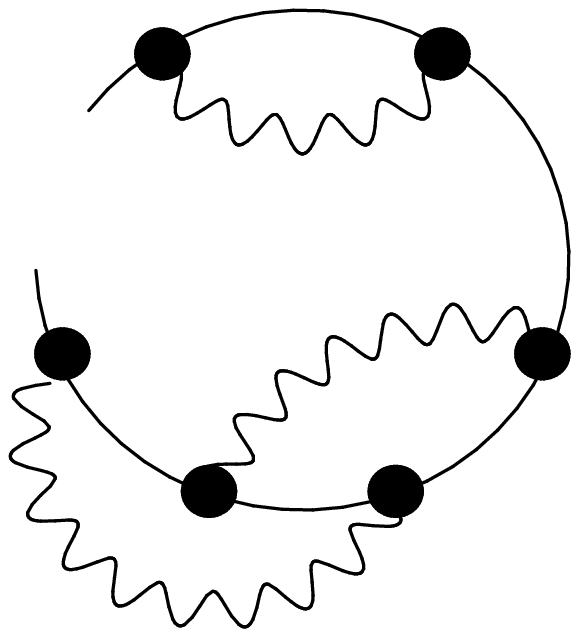}}
\put(3.5,0.3)

\put(8.8,0.6){$+$}

\put(9.1,-0.45){\includegraphics[scale=0.24]{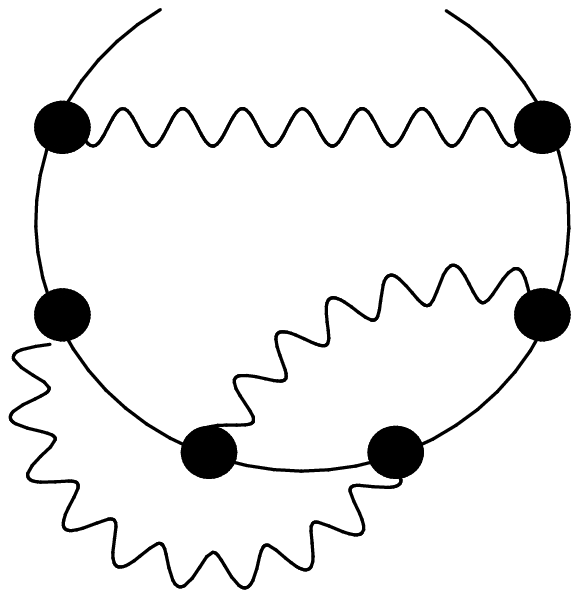}}
\put(3.5,0.3)

\put(11.2,0.6){$+\ \ \ldots$}

\end{picture}
\bigskip

\noindent For diagrams, which can not be made disconnected by two
cuts of the matter loop, this gives the corresponding contribution
to the anomalous dimension.

If a method, based on substituting a solution of the Ward identity
into the Schwinger--Dyson equation, is used, a sum of the
considered diagrams is contained in the effective diagrams that
can be expressed in terms of the two-point function of the matter
superfield. These diagrams give

\begin{equation}\label{Gamma_In_Beta}
\beta(\alpha) \leftarrow - \frac{\alpha^2}{\pi}\gamma(\alpha).
\end{equation}

\noindent (A part of this contribution, corresponding to higher
terms in the expansion of $\ln Z$, comes from diagrams, considered
below.)

Now let us calculate the remaining diagrams

\vspace*{1.5cm}
\begin{picture}(0,0)
\put(2.5,0.2){\includegraphics[scale=0.4]{ins3.eps}}
\put(2.3,0.3){$\theta^a\theta_a \bar\theta^b$}
\put(5.0,0.3){$\theta^c\theta_c\bar\theta^d$}

\put(7,0.75){$+$} \put(7.5,0.75){${\displaystyle\frac{1}{2}}$}

\put(8.7,0.2){\includegraphics[scale=0.4]{ins3.eps}}
\put(8.7,0.3){$\theta^a \bar\theta^b$}
\put(11.2,0.3){$\theta^c\theta_c\bar\theta^d$}

\end{picture}

\noindent The considered sum is given by

\begin{eqnarray}\label{Second_Sum}
&& \frac{d}{d\ln\Lambda} \mbox{Tr}\Big(\theta^4 \Big(\frac{1}{32}
(\gamma^\mu)_d{}^c \Big[y_\mu^*,\frac{\bar D^2 D_c}{\partial^2} *
\frac{\bar D^d D^2}{\partial^2} *\Big] +\frac{1}{16}
(\gamma^\mu)_d{}^c \frac{\bar D^2 D^2
\partial_\mu}{16\partial^4} * \frac{\bar D^2
D_c}{\partial^2} * \frac{\bar D^d D^2}{\partial^2} *\nonumber\\
&& + \frac{1}{16} (\gamma^\mu)_d{}^c \frac{\bar D^2 D^2
\partial_\mu}{16\partial^4} *\frac{\bar D^d D^2}{\partial^2} *
\frac{\bar D^2 D_c}{\partial^2} * - 4i\Big(\frac{\bar
D^2}{2\partial^2}
*   \frac{\bar D^d D^2}{8\partial^2} *
\frac{\bar D_d D^2}{8\partial^2} * + \frac{\bar D_d
D^c}{2\partial^2} * \frac{\bar D^2 D_c}{8\partial^2} * \nonumber\\
&& \times\frac{\bar D^d D^2}{8\partial^2} * + \frac{\bar D_d
D^2}{4\partial^2}
* \frac{\bar D^2 D^c}{4\partial^2} * \frac{\bar D^2
D_c}{8\partial^2} * \frac{\bar D^d D^2}{8\partial^2} * \Big)\Big).
\end{eqnarray}

The first term in this expression is an integral of a total
derivative. The other terms have not yet been written as an
integral of a total derivative.\footnote{In this case the method,
based on substituting a solution of the Ward identity into
Schwinger--Dyson equation, does not work, because here transversal
parts of the Green functions are essential.} Let us qualitatively
explain, how this can be made. For this purpose we cut the diagram
by any chain of internal lines:

\vspace*{1.5cm}

\begin{picture}(0,0)
\put(6.5,0){$\includegraphics[scale=0.2]{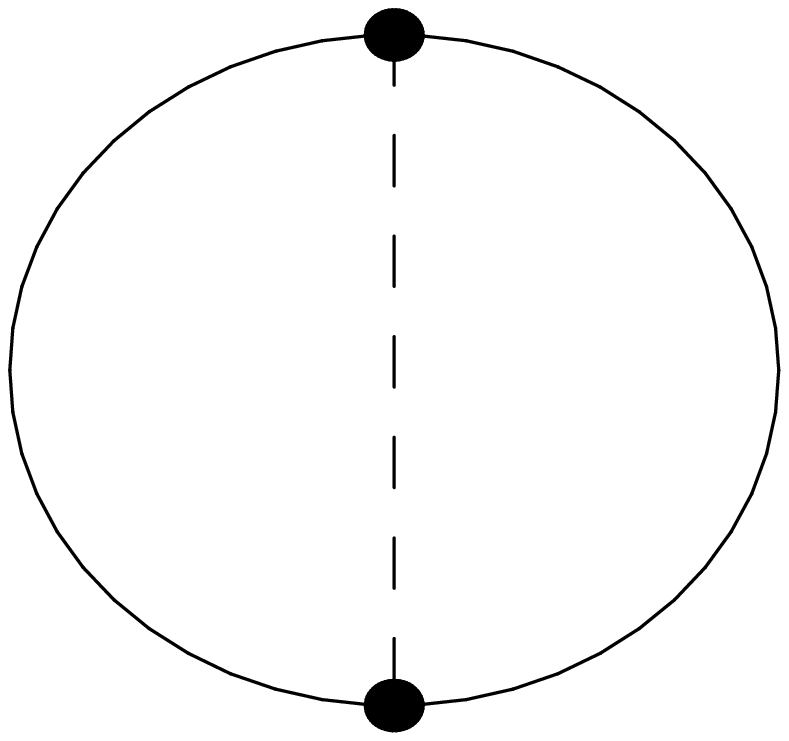}$}
\end{picture}

\noindent Then according to Eq. (\ref{Second_Sum}) (we omit the
first term, which is the integral of a total derivative) the
following variants are possible:

1. There is

\begin{equation}
*\frac{\bar D^a D^2}{8\partial^2}* = [*,\bar\theta^a]
\quad\mbox{or}\quad *\frac{D^2 D^a}{8\partial^2}* = [*,\theta^a]
\end{equation}

\noindent on the left (right) side.

2. There is $ {\displaystyle * \frac{\bar D^a D^2}{8\partial^2}
* \frac{\bar D_a D^2}{8\partial^2} * }$ on the left (right) side.
Then on the other side we have

\begin{equation}
8i \{\theta^b,[\theta_b,*]\} = 8i\Big(- *\frac{\bar
D^2}{4\partial^2}
* + * \frac{\bar D^2 D^b}{4\partial^2} *
\frac{\bar D^2 D_b}{8\partial^2} *\Big).
\end{equation}

3. There is $ {\displaystyle * \frac{\bar D^a D^2}{8\partial^2}
* \frac{\bar D^2 D_b}{8\partial^2} *} $ on the left (right) side.
Then on the other side we obtain

\begin{equation}
2 (\gamma^\mu)_a{}^b \theta^4 [y_\mu^*,*] = 2(\gamma^\mu)_a{}^b
\theta^4 \Big(* \frac{\bar D^2 D^2 \partial_\mu}{8\partial^4} * -
i(\gamma_\mu)_c{}^d * \frac{\bar D^2 D^c}{4\partial^2} *
\frac{\bar D_d D^2}{8\partial^2} *\Big).
\end{equation}

\noindent (The last term in (\ref{Second_Sum}) is contributed by
this expression and by the expression in variant 4. That is why
this expression gives only one half of this term.)

4. There is $ {\displaystyle * \frac{\bar D^2 D_b}{8\partial^2}
* \frac{\bar D^a D^2}{8\partial^2} * } $ on the left (right) side.
Then on the other side we have

\begin{equation}
2 (\gamma^\mu)_a{}^b  [y_\mu^*,*] \theta^4 = 2(\gamma^\mu)_a{}^b
\Big(* \frac{\bar D^2 D^2
\partial_\mu}{8\partial^4} * - i(\gamma_\mu)_d{}^c * \frac{\bar D_c
D^d}{2\partial^2} * -i (\gamma_\mu)_d{}^c * \frac{\bar D_c
D^2}{4\partial^2} * \frac{\bar D^2 D^d}{8\partial^2}
*\Big)\theta^4.
\end{equation}

5. All modified propagators are on the same side of the diagram.

Graphically the commutators (or anticommutators), which are
obtained on one side of the diagram, correspond to one of
subdiagrams

\vspace*{1.5cm}

\begin{picture}(0,0)
\put(5.8,0){$\includegraphics[scale=0.2]{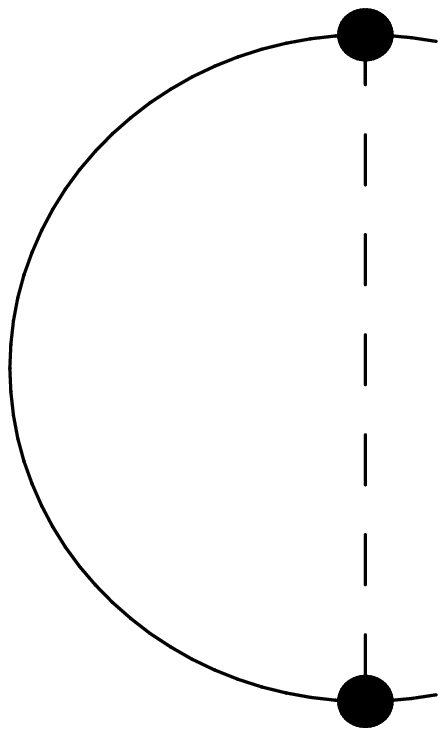}$}

\put(7.3,0.8){or}

\put(7.8,0){$\includegraphics[scale=0.2]{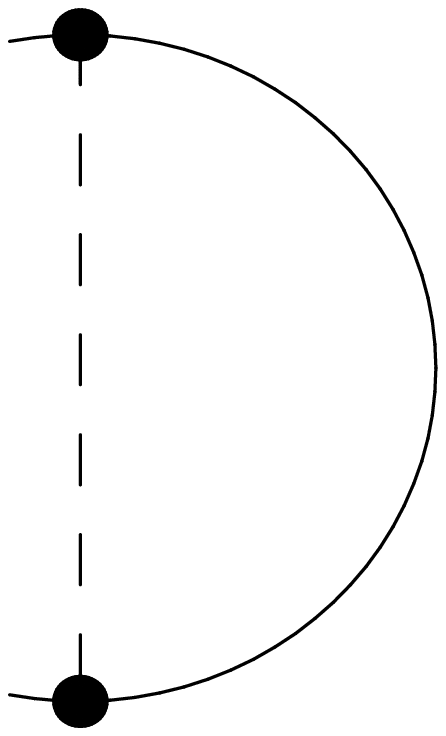}$}

\end{picture}

\noindent with $\theta$-s or $y_\mu^*$ attached to the vertexes.
This subdiagrams can be calculated by a standard supergraph
technique. Using this method $\theta$-parts of the graphs can be
presented as effective lines with all external lines attached to
one of the vertexes. (For this purpose it is necessary to
calculate $\theta$-integrals.)

\vspace*{1.5cm}
\begin{picture}(0,0)
\put(4,0){$\includegraphics[scale=0.2]{half2.eps}$}

\put(5.4,0.9){\vector(1,0){1.4}}

\put(7,0){$\includegraphics[scale=0.2]{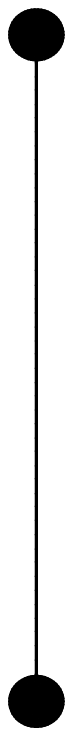}$}

\put(9.4,0.9){\vector(-1,0){1.4}}

\put(9.5,0){$\includegraphics[scale=0.2]{half1.eps}$}

\end{picture}

\noindent Then a number of loops in the $\theta$-part of the
considered diagram is evidently decreased. The procedure can be
repeated, if one perform the following substitutions:

\begin{eqnarray}
&& \frac{\bar D^2 D^2}{16\partial^2} \to A, \quad (P_A=0\quad
\mbox{or}\quad P_A=1);\nonumber\\
&& \frac{\bar D^2 D^a}{8\partial^2} \to -[\theta^a,A\};\qquad\quad
\frac{\bar D^a D^2}{8\partial^2} \to
-[\bar\theta^a,A\};\nonumber\\
&& -\frac{\bar D^2}{4\partial^2} \to
[\theta^a,[\theta_a,A\}\};\qquad \theta^4 \frac{\bar D^2 D^2
\partial_\mu}{8\partial^4} \to \theta^4 [y_\mu^*, A];\nonumber\\
&& \Big(\frac{\bar D^2 D^2
\partial_\mu}{8\partial^4} - i(\gamma^\mu)_d{}^c
\frac{\bar D_c D^d}{2\partial^2}\Big) \theta^4 \to [y_\mu^*,
A]\theta^4.
\end{eqnarray}

\noindent (Also it is necessary to include the factors $(-1)^P$,
but we will not discuss this here. Using another method, based on
this arguments, this signs can be easily taken into account.) The
procedure can be repeated using the identities

\begin{eqnarray}
&& [\theta_a, AB\} = [\theta_a,A\} B + (-1)^{P_A} A
[\theta_a,B\};\vphantom{\Big(}\nonumber\\
&& [y_\mu^*, AB] \theta^4 = [y_\mu^*, A] \theta^4 B + A [y_\mu^*,
B] \theta^4 -2i (-1)^{P_A} (\gamma_\mu)_b{}^a
[\bar\theta_a,A\}[\theta^b,B\}\theta^4 + O(\theta^3);
\vphantom{\Big(}\nonumber\\
&& \theta^4 [y_\mu^*, AB] = \theta^4 [y_\mu^*, A] B + A \theta^4
[y_\mu^*, B] - 2i (-1)^{P_A} (\gamma_\mu)_b{}^a \theta^4
[\theta^b,A\} [\bar\theta_a,
B\} + O(\theta^3);\vphantom{\Big(}\nonumber\\
&& [\theta^a [\theta_a, AB\}\} = [\theta^a [\theta_a, A\}\} B + A
[\theta^a [\theta_a, B\}\} + 2 (-1)^{P_A} [\theta^a, A\}
[\theta_a, B\}.\vphantom{\Big(}
\end{eqnarray}

It is easy to see that at the last step we can obtain no more than
three subdiagrams. Let us denote expressions for these subdiagrams
by $A$, $B$, and $C$. They can be written as differential
operators, constructed from the supersymmetric covariant
derivatives $D$, $\bar D$ and the usual derivatives
$\partial_\mu$. (It is important that they do not explicitly
depend on $\theta$.) Then according to Eq. (\ref{Second_Sum})
finally we obtain

\begin{eqnarray}
&& -2\frac{d}{d\ln\Lambda}\mbox{Tr}\Big(\theta^4 \Big(
(\gamma^\mu)^{ab} [y_\mu^*,A] [\bar\theta_b, B\}[\theta_a, C\} +
(\gamma_\mu)^{ab} (-1)^{P_A}
[\theta_a,B\} [\bar\theta_b, C\} [y_\mu^*,A] \vphantom{\frac{1}{2}}\qquad
\nonumber\\
&& -4i [\theta^a,[\theta_a, A\}\}
[\bar\theta^b,B\}[\bar\theta_b,C\}\Big)\Big) +\mbox{cyclic perm.
of $A$, $B$, $C$}.\qquad\vphantom{\frac{1}{2}}
\end{eqnarray}

\noindent In the Appendix we prove the identity

\begin{eqnarray}\label{Triple_Identity}
&& \mbox{Tr}\Big(\theta^4 \Big( (\gamma^\mu)^{ab} [y_\mu^*,A]
[\bar\theta_b, B\}[\theta_a, C\} + (\gamma_\mu)^{ab} (-1)^{P_A}
[\theta_a,B\} [\bar\theta_b, C\} [y_\mu^*,A] \vphantom{\frac{1}{2}}\nonumber\\
&&\qquad\qquad\qquad\qquad -4i [\theta^a,[\theta_a, A\}\}
[\bar\theta^b,B\}[\bar\theta_b,C\}\Big)\Big) +\mbox{cyclic
perm. of $A$, $B$, $C$}\qquad\vphantom{\frac{1}{2}}\nonumber\\
&& = \frac{1}{3}\mbox{Tr}\Big(\theta^4 (\gamma^\mu)^{ab}
\Big[y_\mu^*, A [\bar\theta_b,B\} [\theta_a, C\} + (-1)^{P_A}
[\theta_a,B\}[\bar\theta_b, C\}A \Big]\nonumber\\
&&\qquad\qquad\qquad\qquad\qquad\qquad\qquad\qquad\qquad\qquad\qquad
+\mbox{cyclic perm. of $A$, $B$, $C$}.\qquad\vphantom{\frac{1}{2}}
\end{eqnarray}

\noindent Thus, the sum of all remaining diagrams is given by an
integral of a total derivative. However, this integral is not
equal to 0, because singularities, proportional to
$\delta$-functions, can appear. Taking them into account, it is
possible to obtain Eq. (\ref{Gamma_In_Beta})
\cite{Future_Publication}.

\section{Conclusion}
\hspace{\parindent}

In this paper we investigate a conjecture that in supersymmetric
theories, regularized by higher covariant derivatives, all
integrals, defining the $\beta$-fucntion, are integrals of total
derivatives. This fact allows to calculate at least one of the
integrals analytically, so that the factorization of integrands
into total derivatives is in the close connection with the NSVZ
exact $\beta$-function.

For $N=1$ SQED, regularized by higher derivatives, we presented
qualitative arguments, which demonstrate an origin of the
factorization of the integrands into total derivatives. Although
they are applicable in all orders, the proof, presented in this
paper, was not made by a completely rigorous method. Also, we have
not considered contributions of the Pauli--Villars fields.
Nevertheless, this proof allows to understand, why the
factorization of integrands into total derivatives takes place.
(Another proof was given in Ref. \cite{Smilga}.) The complete
rigorous proof (for $N=1$ SQED), which, in particular, includes
calculation of Pauli--Villars fields contributions will be
described in a future paper.

\bigskip
\bigskip

\noindent {\Large\bf Acknowledgements.}

\bigskip

\noindent This work was supported by RFBR grant No 08-01-00281a. I
am very grateful to A.L.Kataev, A.A.Slavnov, and A.V.Smilga for
valuable discussions.

\appendix

\section{Integrals, defining the three-loop $\beta$-function}
\hspace{\parindent}\label{Appendix_Three_Loop}

Here we present expressions for the three-loop integrals.
Actually, they were obtained in Ref. \cite{3LoopHEP}, but here we
write them in a more beautiful form. The $\beta$-function is given
by the following integrals:

\begin{eqnarray}
\frac{\beta(\alpha_0)}{\alpha_0^2} = \frac{d}{d\ln\Lambda}
\Big(d^{-1}(\alpha_0,\Lambda/p) - \alpha_0^{-1}\Big)\Big|_{p=0} =
16\pi (A_1 + A_2 + A_3),
\end{eqnarray}

\noindent where\footnote{This result was also presented in Ref.
\cite{Review}, but some sign in $A_3$ were written incorrectly.}

\begin{eqnarray}\label{Three_Loop_Beta1}
&& A_1 = -\frac{1}{2}\sum\limits_I c_I \int \frac{d^4q}{(2\pi)^4}
\frac{1}{q^2} \frac{d}{dq^2}
\frac{d}{d\ln\Lambda}\Big(\ln(q^2+M_I^2) +
\frac{M_I^2}{q^2+M_I^2}\Big);\\
\vphantom{0}\nonumber\\
\label{Three_Loop_Beta2} && A_2 = -2e^2 \int \frac{d^4q}{(2\pi)^4}
\frac{1}{q^2} \frac{d}{dq^2} \frac{d}{d\ln\Lambda} \int
\frac{d^4k}{(2\pi)^4} \frac{1}{k^2 R_k^2}\Bigg(\frac{1}{(k+q)^2} -
\sum\limits_{J} c_J \frac{q^4}{(q^2+M_J^2)^2}
\nonumber\\
&& \times\frac{1}{((k+q)^2+M_J^2)} \Bigg) \Bigg[ R_k
\Big(1+\frac{e^2}{4\pi^2} \ln\frac{\Lambda}{\mu}\Big) - \int
\frac{d^4t}{(2\pi)^4}\,\frac{2 e^2}{t^2 (k+t)^2}
\nonumber\\
\label{Three_Loop_Beta3} &&  + \sum\limits_I c_I \int
\frac{d^4t}{(2\pi)^4}\,
\frac{2 e^2}{(t^2+M_I^2) ((k+t)^2+M_I^2)} \Bigg];\\
\vphantom{0}\nonumber\\
&& A_3 = \int \frac{d^4q}{(2\pi)^4}\,\frac{1}{q^2}\frac{d}{dq^2}
\frac{d}{d\ln\Lambda}\int \frac{d^4k}{(2\pi)^4}
\frac{d^4l}{(2\pi)^4} \frac{4 e^4}{k^2 R_k\, l^2 R_l}
\Bigg\{\frac{1}{(q+k)^2}\Bigg[- \frac{1}{2 (q+l)^2} \nonumber\\
&& + \frac{2 k^2}{(q+k+l)^2 (q+l)^2} - \frac{1}{(q+k+l)^2} \Bigg]
-\sum\limits_{I} c_I
\frac{q^4}{(q^2+M_I^2)^2} \frac{1}{((q+k)^2+M_I^2)}\times\nonumber\\
&&\times \Bigg[- \frac{1}{2 ((q+l)^2+M_I^2) } + \frac{2
k^2}{((q+k+l)^2+M_I^2) ((q+l)^2+M_I^2)}
\nonumber\\
&& - \frac{1}{((q+k+l)^2+M_I^2) } + \frac{2
M_I^2}{((q+k)^2+M_I^2)((q+k+l)^2+M_I^2)
} \nonumber\\
&&  +\frac{2 M_I^2}{(q^2+M_I^2) ((q+l)^2+M_I^2) } +
\frac{2M_I^2}{((q+l)^2+M_I^2) ((q+k+l)^2+M_I^2)}\Bigg] \Bigg\}.
\end{eqnarray}

\noindent Here for simplicity we denote $R_k \equiv
R(k^2/\Lambda^2)$. $A_1$ is a one-loop result. $A_2$ is a sum of
two-loop diagrams, three-loop diagrams with two loops of the
matter superfields, and diagrams with insertions of counterterms,
arising from renormalization of the coupling constant. $A_3$ is a
sum of three-loop diagrams with a single loop of the matter
superfields.

The two-point Green function of the matter superfield in the
two-loop approximation is given by the following integrals:

\begin{eqnarray}\label{Two_loop_LnG}
&& \ln G = -\int \frac{d^4k}{(2\pi)^4} \frac{2 e_0^2}{k^2 R_k
(k+q)^2}\Bigg[ 1 - \frac{1}{R_k}\int
\frac{d^4t}{(2\pi)^4}\,\frac{2 e_0^2}{t^2 (k+t)^2} + \sum\limits_I
c_I \frac{1}{R_k}\int \frac{d^4t}{(2\pi)^4}\,
\nonumber\\
&& \times \frac{2 e_0^2}{(t^2+M_I^2) ((k+t)^2+M_I^2)} \Bigg] +
\int \frac{d^4k}{(2\pi)^4} \frac{d^4l}{(2\pi)^4} \frac{e_0^4}{k^2
R_k
l^2 R_l} \Bigg( -\frac{2}{(q+k)^2 (q+l)^2} \nonumber\\
&& -\frac{4}{(q+k)^2(q+k+l)^2} + \frac{8k^2-4q^2}{(q+k)^2
(q+k+l)^2 (q+l)^2} \Bigg).
\end{eqnarray}

\noindent Therefore, taking into account one-loop renormalization
of the coupling constant, we obtain

\begin{eqnarray}\label{Two_Loop_Gamma}
&&\hspace*{-5mm} \gamma(\alpha_0) = - 2e^2 \int
\frac{d^4k}{(2\pi)^4} \frac{d}{d\ln\Lambda} \frac{1}{k^4 R_k^2}
\Bigg[R_k \Big(1+\frac{e^2}{4\pi^2}\ln\frac{\Lambda}{\mu} \Big) -
\int
\frac{d^4t}{(2\pi)^4}\,\frac{2 e^2}{t^2 (k+t)^2} +\nonumber\\
&&\hspace*{-5mm} + \sum\limits_I c_I \int \frac{d^4t}{(2\pi)^4}\,
\frac{2 e^2}{(t^2+M_I^2) ((k+t)^2+M_I^2)} \Bigg] - \int
\frac{d^4k}{(2\pi)^4} \frac{d^4l}{(2\pi)^4}\,\frac{d}{d\ln
\Lambda}\frac{4 e^4 k_\mu l_\mu}{k^4 R_k\,l^4 R_l
(k+l)^2}.\nonumber\\
\end{eqnarray}

\noindent This expression is finite both in the UV and IR regions.
The UV finiteness is ensured by the regularization. The integral
is IR finite due to the differentiation with respect to
$\ln\Lambda$, which should be performed before the integration.

In order to obtain Eq. (\ref{Relation}) it is necessary to use the
identity

\begin{equation}\label{Integral_Of_Total_Derivative}
\int \frac{d^4q}{(2\pi)^4}\frac{1}{q^2} \frac{d}{dq^2} f(q^2) =
\frac{1}{16\pi^2} \Big(f(q^2=\infty) - f(q^2=0)\Big).
\end{equation}

\section{Proof of identity (\ref{Triple_Identity})}
\hspace{\parindent}\label{Appendix_Identity}

Let us consider

\begin{eqnarray}
&& X\equiv \mbox{Tr}\Big\{\theta^4 \Big( (\gamma^\mu)^{ab}
[y_\mu^*,A] [\bar\theta_b, B\}[\theta_a, C\} + (\gamma^\mu)^{ab}
(-1)^{P_A}
[\theta_a,B\} [\bar\theta_b, C\} [y_\mu^*,A]\nonumber\\
&& -4i [\theta^a,[\theta_a,
A\}\}[\bar\theta^b,B\}[\bar\theta_b,C\}\Big)\Big\} +\mbox{cyclic
perm. of $A$, $B$, $C$},
\end{eqnarray}

\noindent where $A$, $B$, and $C$ are differential operators,
constructed from the supersymmetric covariant derivatives. It is
important that they do not explicitly depend on $\theta$.
Certainly, we assume that

\begin{equation}
P_A + P_B + P_C = 0(\mbox{mod}\ 2).
\end{equation}

\noindent Using the identities

\begin{eqnarray}
&& \mbox{Tr} \Big([y_\mu^*,A] B\Big) = - \mbox{Tr}\Big(A[y_\mu^*,B]\Big);\nonumber\\
&& (-1)^{P_B} [[A,B\},C\} + (-1)^{P_A}[[C,A\},B\} +
(-1)^{P_C}[[B,C\},A\} =0,
\end{eqnarray}

\noindent we obtain

\begin{eqnarray}
&& X = (\gamma^\mu)^{ab} \mbox{Tr} \Big\{\theta^4
\Big(\Big[y_\mu^*, A [\bar\theta_b, B\}[\theta_a, C\} + (-1)^{P_A}
[\theta_a,B\} [\bar\theta_b, C\} A\Big]
- A [\bar\theta_b,[y_\mu^*,B]\}\qquad \nonumber\\
&& \times [\theta_a, C\} - A [\bar\theta_b, B\}[\theta_a,
[y_\mu^*, C]\} - (-1)^{P_A}[\theta_a, [y_\mu^*,B]\} [\bar\theta_b,
C\}A - (-1)^{P_A}[\theta_a,B\}\qquad
\nonumber\\
&& [\bar\theta_b, [y_\mu^*, C]\}A\Big)\Big\} -4i \mbox{Tr}
\Big(\theta^4 [\theta^a,[\theta_a,
A\}\}[\bar\theta^b,B\}[\bar\theta_b,C\}\Big) +\mbox{cyclic perm.
of $A$, $B$, $C$}.\qquad
\end{eqnarray}

\noindent The similar operation is repeated for $\theta$-s in
double commutators:

\begin{eqnarray}\label{X1}
&& X = (\gamma^\mu)^{ab} \mbox{Tr} \Big\{\theta^4
\Big(\Big[y_\mu^*, A [\bar\theta_b, B\}[\theta_a, C\} + (-1)^{P_A}
[\theta_a,B\} [\bar\theta_b, C\} A\Big]+ (-1)^{P_A} [\bar\theta_b,
A\}
\nonumber\\
&& \times [y_\mu^*,B] [\theta_a, C\} + (-1)^{P_B} A [y_\mu^*,B]
[\bar\theta_b, [\theta_a, C\}\} - (-1)^{P_C} [\theta_a, A\}
[\bar\theta_b, B\} [y_\mu^*, C]
\vphantom{\Big(}\nonumber\\
&& - (-1)^{P_B} A [\theta_a, [\bar\theta_b, B\}\} [y_\mu^*, C] -
[y_\mu^*,B] [\bar\theta_b, C\}[\theta_a, A\} + (-1)^{P_C}
[y_\mu^*,B] [\theta_a,[\bar\theta_b, C\}\}A
\vphantom{\Big(}\nonumber\\
&& - (-1)^{P_C}[\bar\theta_b,[\theta_a,B\}\} [y_\mu^*, C]A +
(-1)^{P_B}[\theta_a,B\} [y_\mu^*, C] [\bar\theta_b, A\} \Big)\Big)
\vphantom{\Big(}\nonumber\\
&& -4i \mbox{Tr} \Big(\theta^4 [\theta^a,[\theta_a,
A\}\}[\bar\theta^b,B\}[\bar\theta_b,C\}\Big\} +\mbox{cyclic perm.
of $A$, $B$, $C$}.
\end{eqnarray}

\noindent Commutators with $y_\mu^*$ are proportional to the first
degree of $\theta$:

\begin{equation}
[y_\mu^*,A] = -2i (\gamma^\mu)^{ab}\theta_a [\bar\theta_b,A\} +
O(\theta^0).
\end{equation}

\noindent Thus, it is necessary to be careful, commuting
$\theta^4$ with $A$, $B$ and $C$. For example, taking into account
that all expressions, containing $(\theta)^n$ with $n<4$, vanish,
we obtain

\begin{eqnarray}
&& (-1)^{P_A} (\gamma^\mu)^{ab}\mbox{Tr}\,\Big(\theta^4
[\bar\theta_b, A\} [y_\mu^*,B] [\theta_a, C\}\Big) =
(-1)^{P_B+1}(\gamma^\mu)^{ab}\mbox{Tr}\Big(\theta^4 [\theta_a, C\}
[\bar\theta_b, A\} [y_\mu^*,B]
\quad\nonumber\\
&& - 2\bar\theta^c\bar\theta_c \theta^d [\theta_d,[\theta_a,C\}\}
[\bar\theta_b, A\}
(-2i)(\gamma_\mu)^{ef} \theta_e [\bar\theta_f,B\}\Big)\qquad\nonumber\\
&& = \mbox{Tr}\Big((-1)^{P_B+1} \theta^4 (\gamma^\mu)^{ab}
[\theta_a, C\} [\bar\theta_b, A\} [y_\mu^*,B] + 4i \theta^4
[\theta^a,[\theta_a,C\}\} [\bar\theta^b, A\}
[\bar\theta_b,B\}\Big).
\end{eqnarray}

\noindent Similarly one can derive the following identities:

\begin{eqnarray}
&& (-1)^{P_B} (\gamma^\mu)^{ab}
\mbox{Tr}\Big(\theta^4[\theta_a,B\}[y_\mu^*,C][\bar\theta_b,A\}\Big)
\\
&&\qquad =
\mbox{Tr}\Big(-\theta^4(\gamma^\mu)^{ab}[y_\mu^*,C][\bar\theta_b,A\}[\theta_a,B\}
+4i\theta^4 [\theta^a,[\theta_a,B\}\}[\bar\theta^b,C\}[\bar\theta_b,A\}\Big);\nonumber\\
&& (-1)^{P_B+1} (\gamma^\mu)^{ab} \mbox{Tr}\Big(\theta^4 A
[\theta_a,[\bar\theta_b,B\}\}[y_\mu^*,C]\Big)
\\
&&\qquad =\mbox{Tr}\Big((-1)^{P_C+1}\theta^4
(\gamma^\mu)^{ab}[\theta_a,[\bar\theta_b,B\}\}[y_\mu^*,C]A -4i
(-1)^{P_A}\theta^4 [\theta^a, A\}[\theta_a,
[\bar\theta^b, B\}\}[\bar\theta_b,C\}\Big);\nonumber\\
&& (-1)^{P_B} (\gamma^\mu)^{ab} \mbox{Tr}\Big(\theta^4 A
[y_\mu^*,B][\bar\theta_b,[\theta_a, C\}\}\Big)
\\
&&\qquad = (-1)^{P_C}\mbox{Tr}\Big(\theta^4 (\gamma^\mu)^{ab}
[y_\mu^*,B][\bar\theta_b, [\theta_a,C\}\}A + 4i \theta^4
[\theta^a, A\}[\bar\theta_b, B\} [\bar\theta^b, [\theta_a,
C\}\}\Big).\nonumber
\end{eqnarray}

\noindent Using these equations, after some algebra $X$ can be
rewritten as

\begin{eqnarray}
&& X = \mbox{Tr}\Big\{\theta^4 (\gamma^\mu)^{ab}
\Big(\Big[y_\mu^*, A[\bar\theta_b,B\}[\theta_a,C\} +
(-1)^{P_A}[\theta_a,B\}[\bar\theta_b,C\}A \Big] -2 [y_\mu^*, A]
[\bar\theta_b, B\}
\nonumber\\
&& \times [\theta_a, C\} -2 (-1)^{P_A} [\theta_a,
B\}[\bar\theta_b, C\}[y_\mu^*,A]\Big) + 8i \theta^4
[\theta^a,[\theta_a,A\}\} [\bar\theta^b,B\}[\bar\theta_b,C\}
\Big\}
\nonumber\\
&& +\mbox{cyclic perm. of $A$, $B$, $C$}.
\end{eqnarray}

\noindent Comparing this expression with the definition of $X$, we
obtain

\begin{eqnarray}
&& X = -2X + \Bigg(\mbox{Tr}\Big(\theta^4 (\gamma_\mu)^{ab}
\Big[y_\mu^*, A [\bar\theta_b,B\} [\theta_a, C\} + (-1)^{P_A}
[\theta_a,B\}[\bar\theta_b, C\}A \Big]\Big)\nonumber\\
&&+\mbox{cyclic perm. of $A$, $B$, $C$}\smash{\Bigg)}.
\end{eqnarray}

\noindent Therefore,

\begin{eqnarray}
&& X = \frac{1}{3}\mbox{Tr}\Big(\theta^4 (\gamma_\mu)^{ab}
\Big[y_\mu^*, A [\bar\theta_b,B\} [\theta_a, C\} + (-1)^{P_A}
[\theta_a,B\}[\bar\theta_b, C\}A \Big]
\nonumber\\
&&+\mbox{cyclic perm. of $A$, $B$, $C$}.
\end{eqnarray}

\noindent This completes the proof.


\end{document}